\documentclass[%
 reprint,
superscriptaddress,
 amsmath,amssymb,
 aps,
 prl,
]{revtex4-2}

\usepackage{graphicx}% Include figure files
\usepackage{dcolumn}% Align table columns on decimal point
\usepackage{bm}% bold math
\usepackage{xcolor}
\usepackage{comment}
\usepackage{hyperref}% add hypertext capabilities
\begin{document}

\preprint{APS/123-QED}

\title{NV center magnetometry up to 130 GPa as if at ambient pressure} % Force line breaks with \\

% other suggestions: 

% Exploring the Megabar Range with Nitrogen-Vacancy Centers: Achieving Ambient Pressure-Like Optical Magnetometry in the Diamond Anvil Cell
% Optical Magnetometry in the Megabar Range: Mitigating Stress Effects in Nitrogen-Vacancy Centers of the Diamond Anvil Cell
% Microstructured Diamond Anvil Cells for Nitrogen-Vacancy Centers: Achieving Megabar Range Magnetometry through Stress Mitigation
% Nitrogen-Vacancy center ground state behaviour under non-hydrostatic stress probed by optically detected magnetic resonance inside a diamond anvil cell

\author{Antoine Hilberer}
\affiliation{Universit\'e Paris-Saclay, CNRS, ENS Paris-Saclay, CentraleSupelec, LuMIn, F-91190 Gif-sur-Yvette, France}

\author{Loïc Toraille}
%\email{loic.toraille@cea.fr}
\affiliation{CEA DAM DIF, F-91297 Arpajon, France}
\affiliation{Université Paris-Saclay, CEA, Laboratoire Matière en Conditions Extrêmes, 91680 Bruyères-le-Châtel, France}

\author{Cassandra Dailledouze}
\affiliation{Universit\'e Paris-Saclay, CNRS, ENS Paris-Saclay, CentraleSupelec, LuMIn, F-91190 Gif-sur-Yvette, France}

\author{Marie-Pierre Adam}
\affiliation{Universit\'e Paris-Saclay, CNRS, ENS Paris-Saclay, CentraleSupelec, LuMIn, F-91190 Gif-sur-Yvette, France}

\author{Liam Hanlon}
\affiliation{Universit\'e Paris-Saclay, CNRS, ENS Paris-Saclay, CentraleSupelec, LuMIn, F-91190 Gif-sur-Yvette, France}

\author{Gunnar Weck}
\affiliation{CEA DAM DIF, F-91297 Arpajon, France}
\affiliation{Université Paris-Saclay, CEA, Laboratoire Matière en Conditions Extrêmes, 91680 Bruyères-le-Châtel, France}

\author{Martin Schmidt}
\affiliation{Universit\'e Paris-Saclay, CNRS, ENS Paris-Saclay, CentraleSupelec, LuMIn, F-91190 Gif-sur-Yvette, France}

\author{Paul Loubeyre}
\affiliation{CEA DAM DIF, F-91297 Arpajon, France}
\affiliation{Université Paris-Saclay, CEA, Laboratoire Matière en Conditions Extrêmes, 91680 Bruyères-le-Châtel, France}

\author{Jean-François Roch}
\email{jean-francois.roch@ens-paris-saclay.fr}
\affiliation{Universit\'e Paris-Saclay, CNRS, ENS Paris-Saclay, CentraleSupelec, LuMIn, F-91190 Gif-sur-Yvette, France}

\date{\today}% It is always \today, today,
 % but any date may be explicitly specified

\begin{abstract}
Engineering a layer of nitrogen-vacancy (NV) centers on the tip of a diamond anvil creates a multipurpose quantum sensors array for high pressure measurements, especially for probing magnetic and superconducting properties of materials. Expanding this concept above 100~GPa appears to be a substantial challenge. We observe that deviatoric stress on the anvil tip sets a limit at 40-50~GPa for practical magnetic measurements based on optically detected magnetic resonance (ODMR) of NV centers under pressure. We show that this limit can be circumvented up to at least 130~GPa by machining a micropillar on the anvil tip to create a quasi-hydrostatic stress environment for the NV centers. This is quantified using the pressure dependence of the diamond Raman shift, the NV ODMR dependence on applied magnetic field, and NV photoluminescence spectral shift. This paves the way for direct and reliable detection of the Meissner effect in superconductors above 100~GPa, such as super-hydrides.

\end{abstract}

\maketitle

%\tableofcontents

%%%%%%%%%%%%%%%%%%%%%%%%%%%%%%%%%%%%%%%%%%%%%%%%%%%%%%%%%%%\section*{Introduction}

\textit{Introduction.} The diamond anvil cell (DAC) is routinely used to synthesize compounds under megabar (100~GPa) pressures, exhibiting novel phenomena and remarkable properties. Recent examples such as the observation of metal hydrogen~\cite{loubeyre2020}, superconductivity close to ambient temperature in superhydrides~\cite{flores-livas2020, drozdov2019, somayazulu2019}, or superionic water ice~\cite{weck2022} are lacking detailed magnetic or transport measurements for their definite proof and clear understanding. In particular, magnetic measurements remain challenging at megabar pressures because they are mainly based on flux detection by inductive coils and must thus extract the signal of the few-micrometers samples from the much larger magnetic background signal of the bulky DAC apparatus.
This constraint can be circumvented by implementing in the DAC sensing methods that exploit the magnetic sensitivity of nitrogen-vacancy (NV) centers in diamond~\cite{lesik2019, hsieh2019, yip2019, shang2019}. This method offers a tabletop optical microscopy instrumentation, the mapping of the magnetic field in the sample chamber with micrometer spatial resolution and the absence of any sensitivity decrease with the sample size down to the micrometer scale. Another key feature is the easy combination with synchrotron X-ray characterizations to correlate the magnetic or superconducting properties with a well-defined crystallographic structure~\cite{toraille2020}. Yet, the extension of this technique to extreme pressures remains a challenge~\cite{doherty2014}. We investigate here how the existence of a deviatoric stress in the diamond anvil sets effective limits to the magnetic response of NV centers localized at the anvil tip to maximize sample proximity~\cite{hsieh2019,lesik2019}. We then propose and implement a method that overcomes that limit and keeps the full NV quantum sensing capabilities at pressures above 100 GPa.

\begin{figure}
\includegraphics[width=\linewidth]{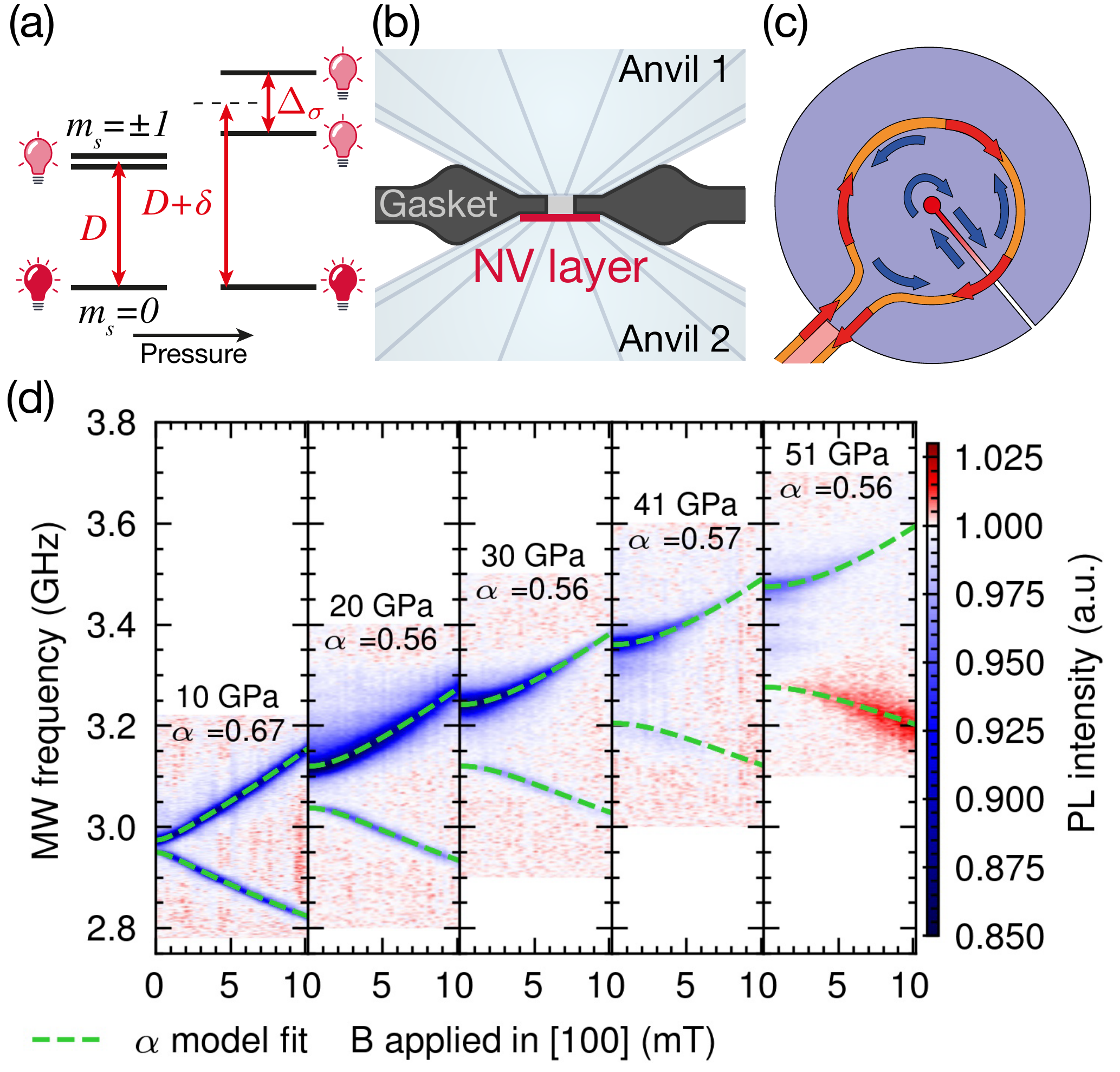}
\caption{\label{Fig1} (a) Energy diagram of the NV center ground state and evolution under stress. (b) Schematic cross-section of the location of NV centers implanted as a layer below the anvil culet surface. (c) Design of the machined gasket compatible with the MW excitation of the NV centers. Red arrows show initial MW excitation current in the wire loop, blue arrows are currents induced into the gasket. The areas shaded in red indicate the intensity of the MW field. (d) ODMR spectra of NV centers implanted in the tip of a standard diamond anvil at different pressures, as a function of a magnetic field applied along the [100] diamond axis. Green dashed lines are fits of the eigenfrequencies computed with the NV ground state Hamiltonian given by eq.~\ref{eq:stress tensor}.}
\end{figure}

\textit{Experimental configuration.} The negatively charged NV center is a point defect of diamond that emits visible photoluminescence (PL) by absorbing green photons and re-emitting red photons (at ambient pressure), with an electronic spin $s=1$ in the ground and excited states. In the absence of external magnetic and stress fields, the $m_s=\pm1$ spin sublevels of the ground state are degenerate and separated by $D=2.87$~GHz from the $m_s=0$ sublevel (Fig.~\ref{Fig1}a). Spin-dependent PL arises from a spin-selective difference in the non-radiative coupling to metastable singlet states, which also induces optical pumping into the $m_s=0$ state under green illumination~\cite{doherty2013}. The energy difference between the sublevels of the ground state can then be read out from the change of the NV luminescence intensity upon scanning the frequency of an additional microwave excitation. Dips in the PL intensity indicate that the excitation microwave frequency is resonant with a transition between two sublevels, leading to optically detected magnetic resonance (ODMR) that can be easily implemented by optically addressing the NV centers through the diamond anvil~\cite{lesik2019}. \\
Here we use the same experimental configuration as in Ref.~\cite{lesik2019}, keeping two crucial characteristics: 1) the NV centers are integrated in the DAC device by mounting a IIas ultra-pure Almax-Boehler design [100]-cut diamond anvil with a dense ensemble of NV centers (typically 10\textsuperscript{4}~NV/\textmu m\textsuperscript{2}) implanted at about 10~nm beneath the anvil surface using a nitrogen Focused Ion Beam (FIB)~\cite{Lesik2013} (Fig.~\ref{Fig1}b);
2) the microwave excitation is applied using an external single-turn coil above the rhenium gasket of the DAC. The metallic gasket is machined with a slit, filled with an epoxy-glue mixture ensuring sample confinement and DAC mechanical stability, that re-distributes the induced currents in the metal, leading to a focusing and amplification of the microwave flux in the sample chamber similarly to a Lenz lens~\cite{meier2017} (Fig.~\ref{Fig1}c).
Upon pressure increase, the PL excitation wavelength was decreased to match the blueshift of the NV absorption spectrum~\cite{doherty2014} by using continuous-wave (cw) lasers at successive wavelengths 532, 488, 457 and 405~nm.
A customized confocal optical microscope was used to collect the PL. 
A static vector magnetic field was applied on the DAC using three Helmholtz coil pairs with an amplitude ranging between 0 and 10~mT. The magnetic field was aligned along the DAC axis with accuracy $\pm0.5^{\circ}$. This orientation corresponds to the diamond [100] crystal axis for which all NV centers have equivalent responses to stress and magnetic field.
Pressure in the DAC was measured using the calibrated diamond Raman phonon mode at the anvil tip~\cite{akahama2004}.\\

\textit{Stress effect on the NV magnetic response.} We performed cw-ODMR experiments on the NV centers under pressures ranging from 10~GPa to 70~GPa. At each pressure point, we collected the ODMR spectrum for the ensemble of NV centers under varying amplitude of the applied magnetic field.

\begin{figure}
\includegraphics[width=\linewidth]{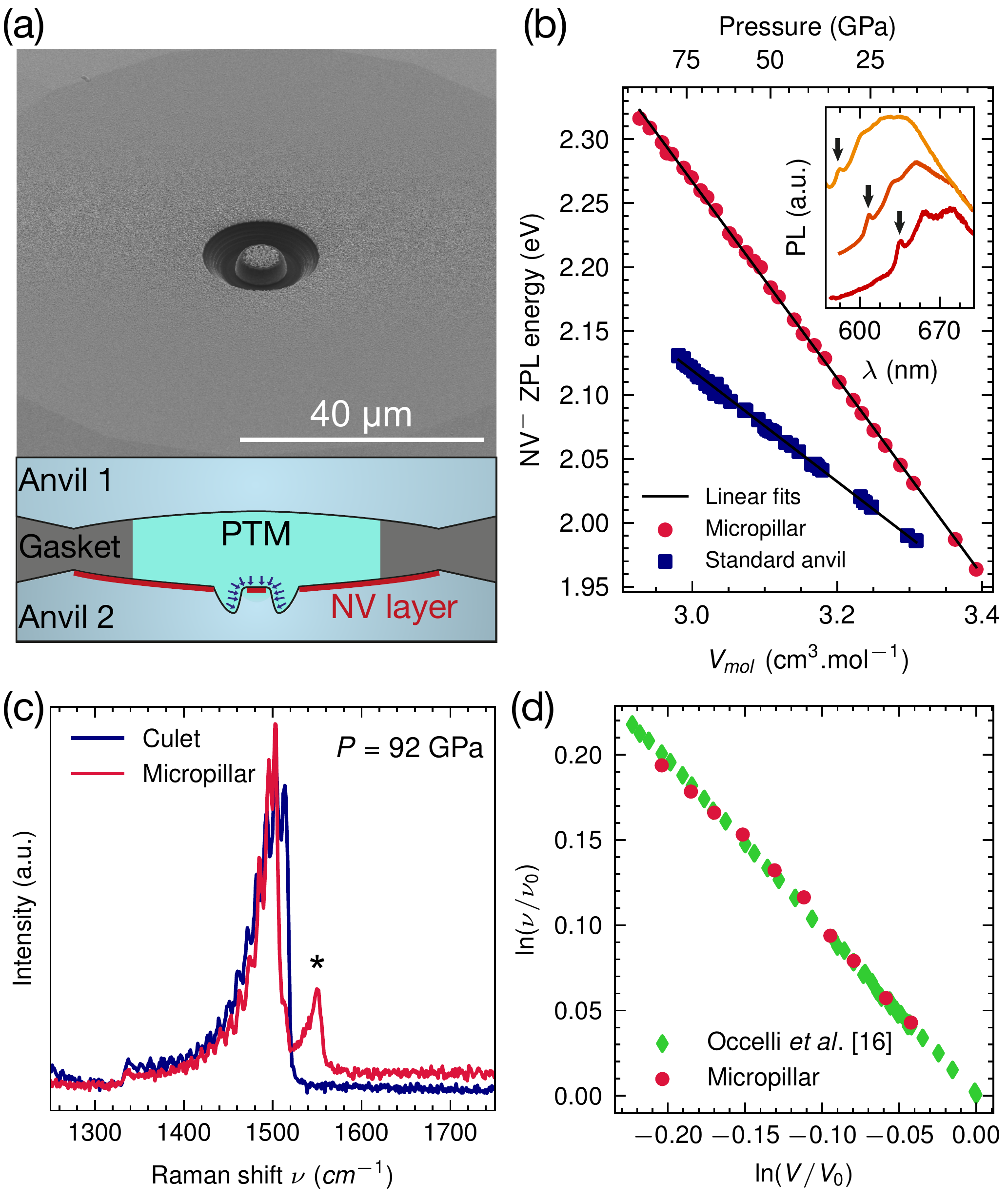}
\caption{\label{Fig2}(a) Scanning electron microscope image of a FIB-machined micropillar on a diamond anvil culet of 100~\textmu m diameter. The bottom panel shows a schematic cross-section with the distortion under pressure of the culet. (b) Energy of the NV center zero-phonon line (ZPL) as a function of pressure and diamond volume, recorded for NV centers implanted in and out of the micropillar. Inset: typical PL spectra of the NV centers recorded at 0, 37 and 78~GPa (bottom to top). The arrows indicate the ZPL position. (c) Diamond Raman spectra recorded on a pressurized microstructured diamond anvil at 92~GPa, on and outside the micropillar. In the spectrum taken on the micropillar, the peak indicated by the star reveals hydrostatic compression. (d) Raman frequency shift measured on the micropillar as a function of relative diamond volume. Data from~\cite{occelli2003} is a reference of the Raman shift of diamond under hydrostatic pressure.}
\end{figure}

%%%%%%%%%%%%%%%%%%%%%%%%%%%%%%%%%%%%%%%%%%%%%%%%%%%%%%%%%%%\section*{Unmodified anvil}

The data are shown in Fig.~\ref{Fig1}d. Four effects of stress on the ODMR signals are observed. First, the zero-field center frequency $D=2.87$~GHz increases almost linearly with a slope of 9.6~MHz/GPa to a value $D+\delta$, where $\delta$ is the pressure induced variation. Second, a splitting $\Delta_\sigma$ appears between the transition lines in the absence of an external magnetic field. This splitting increases almost linearly with pressure with a slope of 3.9~MHz/GPa and originates in deviatoric stress at the anvil culet. 
Consequently, at a given pressure, the quasi-linear evolution of the Zeeman splitting due to the applied magnetic field can only be recovered above a compensating amplitude of the magnetic field that increases with pressure. This detrimental influence of stress hence weakens the NV sensing magnetic sensitivity. Furthermore, the required larger applied bias magnetic field isn't aligned with a given NV axis here, to overlap responses from all NV orientations, and thus mixes the sublevels of the ground state. This mixing perturbs the optically induced spin polarization and quenches the PL~\cite{tetienne2012}.
Third, the shape of the ODMR spectra differs from the conventional symmetrical pair of peaks. The contrast of the low frequency branch becomes gradually smaller than the high frequency branch. After vanishing at a pressure around 40~GPa, a slightly positive contrast reappears (increase of PL at resonance) above 50~GPa under high enough magnetic field.
Finally, the overall observed ODMR contrast decreases severely under pressure.\\

 In the diamond lattice under mechanical stress (or equivalently strain), the Hamiltonian describing the NV center ground state is modified by a spin-mechanical interaction~\cite{hughes1967, barfuss2019} related to the stress tensor $\stackrel{\leftrightarrow}{\sigma}$. The stress tensor must exhibit the cylindrical symmetry of the anvil. At the anvil tip, the stress components parallel ($\sigma_{\parallel}$) and perpendicular ($\sigma_{\perp}$) to the surface differ. 
 Due to continuity of the normal stress component, $\sigma_{\perp}$ is equal to the experimental pressure $P$ in the DAC chamber. The tangential component, $\sigma_{\parallel}$, is reduced by a factor $\alpha$ compared to $\sigma_{\perp}$. Using a simplified model of a semi-infinite anvil with a flat face and a circularly symmetric distribution of pressure applied to this face, the $\alpha$ parameter was estimated about 0.6 ~\cite{ruoff1991}. 
 Neglecting off-diagonal shear stress components, the stress tensor then reads as:
\begin{equation}
\label{eq:stress tensor}
\stackrel{\leftrightarrow}{\sigma}=
\begin{pmatrix}
\alpha P & 0 & 0 \\
0 & \alpha P & 0 \\
0 & 0 & P
\end{pmatrix}.
\end{equation}

Using this stress tensor, the diagonalization of the NV ground state Hamiltonian yields modified spin resonance frequencies which can be approximated to first order as:
\begin{equation}
 {\nu_{\pm} = D + \delta \pm \Delta/2}
 \label{eq:freq:stress}
\end{equation}
where $\delta$ is the spectral shift due to compression, and $\Delta=\sqrt{\Delta_\sigma^2+\Delta_B^2}$ is the quadratic sum of the splittings respectively induced by the stress and by the magnetic field (see Supplementary Material for the full expression). Since eq. (\ref{eq:freq:stress}) is exact only for low off-axis magnetic field, a full numerical diagonalization was used to accurately fit the measured resonance frequencies, as shown by the green dashed lines in fig.~\ref{Fig1}d. Only two parameters, $\alpha$ and $P$, are hence needed to predict the magnetic field response under stress. We obtained a value $\alpha=0.56$ that is essentially constant with pressure, quantifying deviatoric stress close to the 0.6 value given in Ref.~\cite{ruoff1991}.

Deviatoric stress thus introduces major modifications to the NV behavior as the anisotropic compression of the diamond host lattice distorts the $C_{3v}$ symmetry of the NV center. Here we quantified changes within the NV ground triplet states, but the stress dependence of the singlet states and the excited triplet states remains unexplored and is difficult to assess. As a hypothesis, we attribute the observed modification and ultimate loss of ODMR contrast to the effect of deviatoric stress on these levels involved in the contrast mechanism~\cite{goldman2015a}. This hypothesis is corroborated by recent results obtained on microdiamonds compressed quasi-hydrostatically inside the sample chamber of a DAC, for which the ODMR signal could be conserved up to 140~GPa~\cite{dai2022a}. These results converge toward a possible circumventing strategy by ensuring hydrostatic compression of the NV centers. 
\\

\begin{figure*}
\includegraphics[width=\textwidth]{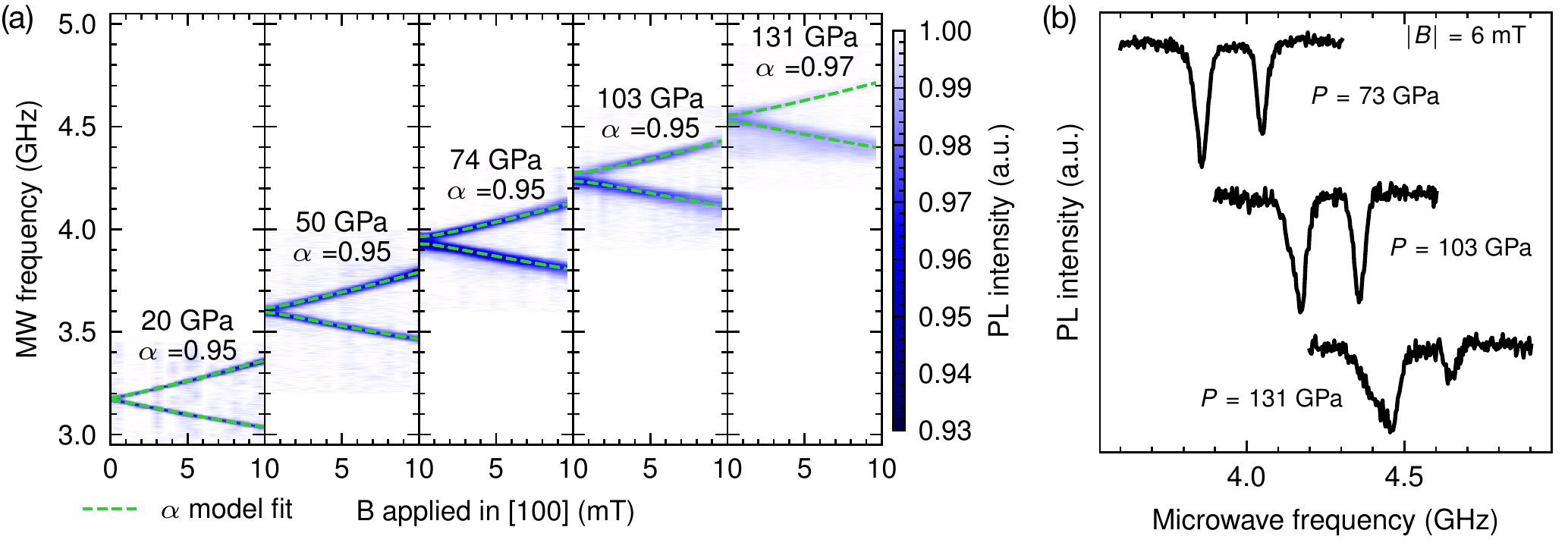}
\caption{\label{Fig3} (a) ODMR spectra obtained from NV centers implanted in a micropillar at varying pressures, as a function of magnetic field applied along the diamond [100] axis. Fitted values of the stress anisotropy parameter $\alpha \simeq 0.95$ indicate quasi-hydrostatic conditions. (b) ODMR spectra recorded for NV centers in the micropillar for a magnetic field of 6~mT amplitude. The signals at 73~GPa, 103~GPa, and 131 GPa are normalized for clarity, with contrast values of 5\%, 3\% and 1.5\% respectively.}
\end{figure*}

\textit{Restoring hydrostaticity with diamond microstructuration.} A strategy to try to mitigate deviatoric stress can be implemented by microstructuring the diamond anvil culet. A successful geometry is presented in Fig.~\ref{Fig2}a. A pillar, 7~\textmu m in diameter and with a 2~\textmu m deep trench around it was FIB-machined on an NV-implanted diamond anvil culet. The pillar surface is thus disconnected from the anvil surface submitted to deviatoric stress induced by anvil cupping tension~\cite{liu2014, li2018}. This also allows the pressure-transmitting medium (PTM) to fill the trench to immerse the pillar in a stress field close to hydrostatic conditions. The pillar is then equivalent to a diamond microdisk that would be integrated in the sample chamber of the DAC but ensures perfect reproducibility and removes any interface with the
diamond culet to optimize PL measurements. As seen below, this design is also very robust and can withstand extreme pressures.

The hydrostaticity of the stress exerted on diamond under pressure can be tested by measuring the Raman frequency of the diamond optical phonon. Under hydrostatic conditions, the dependence of the frequency of the Raman scattering with diamond volume follows a Gruneisen relation of parameter $\gamma = 0.97(1) $ whereas the frequency shift is smaller under deviatoric stress~\cite{occelli2003}. As seen in Fig.~\ref{Fig2}c, the Raman spectra measured at the diamond anvil culet on the micropillar and away from it differ. In both cases, the broad asymmetric peak is associated to the stress distribution within the thickness of the anvil that is optically probed and the high frequency edge is used to estimate the pressure~\cite{akahama2004}. At the micropillar, a well separated peak appears with higher frequency shift. The pressure evolution of its center wavenumber perfectly matches the value obtained for diamond under hydrostatic pressure~\cite{occelli2003} as shown in Fig.~\ref{Fig2}d. This indicates that the tip of the micropillar hosting part of the NV center layer is then close to hydrostatic pressure.

Accordingly the PL spectrum of the NV layer in the micropillar shows a pressure induced blue shift (Fig.~\ref{Fig2}b) that can be quantified with the zero-phonon line (ZPL)~\cite{doherty2014}. While the NV ZPL dependence with pressure is not linear, its evolution becomes linear when plotted versus the compressed diamond volume estimated using the diamond equation of state~\cite{shen2020}. Linear fit gives a slope of $-769 \pm 4$ meV/(cm$^3$·mol$^{-1})$. A similar measurement performed on a non modified diamond anvil yields a weaker slope of $-434 \pm 2$ meV/(cm$^3$·mol$^{-1}$). This significant difference in the pressure dependence of the ZPL is another indication of the deviatoric stress reduction caused by the microstructuration.

\begin{figure} 
\includegraphics[width=\linewidth]{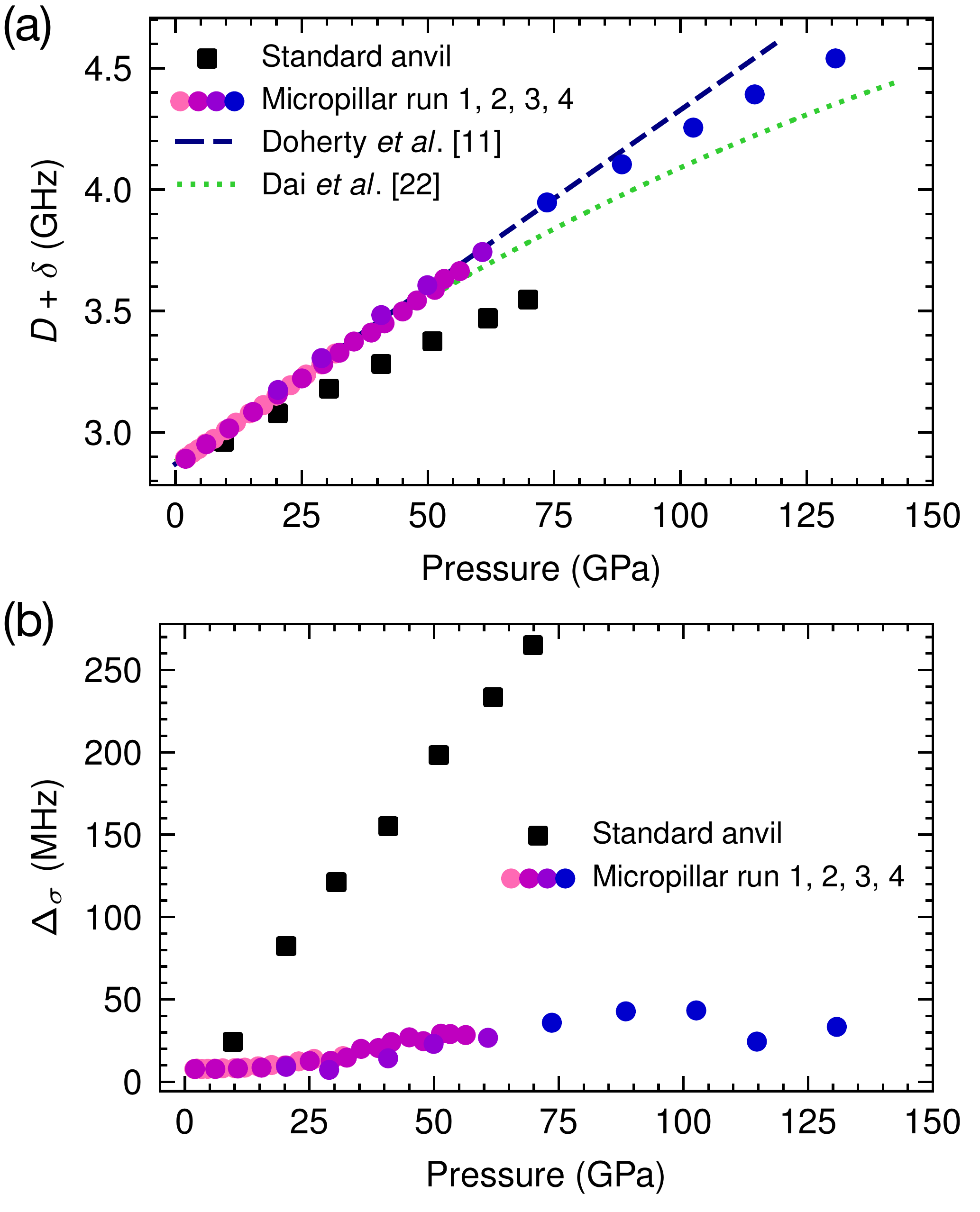}
\caption{\label{Fig4} (a) Pressure dependence of ODMR center frequency $D+\delta$, showing a quasi-linear shift of $13.42 \pm 0.14$~MHz/GPa on the micropillar compared to $9.68 \pm 0.8$~MHz/GPa on the standard anvil. The extrapolation of the values measured up to 60~GPa in~\cite{doherty2014} and the fit up to 140~GPa from~\cite{dai2022a} are given for comparison. 
(b) Pressure dependence of ODMR frequency splitting $\Delta_\sigma$ at zero magnetic field. At the micropillar, $\Delta_\sigma$ increases by $0.29 \pm 0.03$~MHz/GPa instead of $3.89\pm 0.06$~MHz/GPa with the standard geometry of the anvil. }
\end{figure}

ODMR measurements were also performed for the NV centers hosted in the micropillar. As shown in Fig.~\ref{Fig3} corresponding to the pressure evolution up to 130~GPa, most of the detrimental effects previously observed and attributed to deviatoric stress are now suppressed. The spectra consistently show a negative contrast remaining almost constant up to at least 100~GPa. Increasing further the pressure up to 130~GPa (where the experiment was stopped by one of the anvils breaking), a slight decrease of the contrast was observed and is attributed to a degraded efficiency of the microwave excitation for frequencies higher than 4~GHz. The magnetic field response remains also unchanged across the whole tested pressure range. The ODMR spectra exhibit a very low zero-field splitting $\Delta_\sigma$ of $0.29\pm 0.03$~MHz/GPa with increasing pressure, and a shift of the zero-field center frequency $D+\delta$ of $13.42 \pm 0.14$~MHz/GPa. As shown in Fig.~\ref{Fig4} these values differ significantly from those measured for NV centers in standard anvils, and were consistent across four experimental runs performed on different anvils, with pillars machined either using a FIB or a femtosecond laser. Applying the model described above for the spin-mechanical interaction, the evolution of the ODMR eigenfrequencies versus the applied magnetic field were well-fitted using an anisotropy parameter $\alpha \simeq 0.95$ that stays constant within the pressure range tested (Fig.~\ref{Fig3}a). Since $\alpha \simeq 1$ would indicate perfect hydrostaticity, this result gives an independent confirmation of the almost hydrostatic pressure applied on the NV centers in the micropillar.
 Consequently, the microstructuration strategy enables efficient magnetic field sensing at pressures higher than 100~GPa with a sensitivity improved by orders of magnitude compared to the use of a standard anvil with a flat tip (see Supplementary Material).\\

\textit{Conclusion.} Microstructuration of diamond anvils, implemented here by machining a micropillar on the culet, provides quasi-hydrostatic conditions for NV centers implanted in the anvil up to 100~GPa and above. With this design NV magnetic sensing can be implemented under such extreme pressures as if at ambient pressure. This work opens the way to sensitive and spatially resolved magnetic measurements in the constrained environment of the DAC which should now be used for a convincing observation of the Meissner effect in super-hydrides.\\

We are grateful to Olivier Marie and Grégoire Le Caruyer for machining of the diamond culets, to Florent Occelli for assistance in DACs preparation and to Dorothée Colson and Anne Forget for annealing the diamond anvils after nitrogen implantation. This work has received funding from the EMPIR program co-financed by the Participating States and the European Union’s Horizon 2020 research and innovation program (20IND05 QADeT), from the Agence Nationale de la Recherche under the project SADAHPT and the ESR/EquipEx+ program (grant number ANR-21-ESRE-0031), and from the Paris \^Ile-de-France R\'egion in the framework of DIM SIRTEQ. JFR acknowledges support from Institut Universitaire de France.

%\nocite{*}

\bibliography{references}% Produces the bibliography via BibTeX.

\end{document}